\documentclass[useAMS,usenatbib,usegraphicx]{mn2e}

\usepackage{amsmath}

\title[Modelling Spectra using VSTAR]{Modelling the Spectra of Planets, Brown Dwarfs and Stars using VSTAR}
\author[J. Bailey \& L. Kedziora-Chudczer]{Jeremy Bailey,\thanks{E-mail:
j.bailey@unsw.edu.au}
Lucyna Kedziora-Chudczer \\ 
\\
School of Physics, University of New South Wales, NSW 2052, Australia\\
}
\begin{document}

\date{Accepted 2011 September 16; Received 2011 September 15; in original form 2011 August 23}

\pagerange{\pageref{firstpage}--\pageref{lastpage}} \pubyear{2011}

\maketitle

\label{firstpage}

\begin{abstract} We describe a new software package capable of predicting the spectra of
solar-system planets, exoplanets, brown dwarfs and cool stars. The Versatile Software for
Transfer of Atmospheric Radiation (VSTAR) code combines a line-by-line 
approach to molecular and atomic absorption with a full multiple scattering
treatment of radiative transfer. VSTAR is a modular system incorporating an
ionization and chemical equilibrium model, a comprehensive treatment of spectral
line absorption using a database of more than 2.9 billion spectral lines, a
scattering package and a radiative transfer module. We test the methods by comparison 
with other models and benchmark calculations. We present examples of
the use of VSTAR to model the spectra of terrestrial and giant planet in our
own solar system, brown dwarfs and cool stars.
\end{abstract}

\section{Introduction}

Until recently the modelling of the atmospheres of stars \citep[e.g.][]{gray05}
and the modelling of the atmospheres of the Earth and other solar-system planets
 \citep[e.g.][]{liou02} have developed largely independently. Models of stars applied to high
temperature objects with effective temperatures $T_{\mbox{\scriptsize eff}} >$ 3000K, with opacity 
dominated by the line and continuum absorption of atoms and atomic ions, whereas
planetary atmosphere models applied to cool objects $T_{\mbox{\scriptsize eff}} \sim$ 100-300K where the important processes
were molecular absoprtion and scattering from molecules and cloud particles.

This situation changed with the discovery in the mid-1990s of the first
unambiguous brown
dwarf, Gl229B. \citep{nakajima95,oppenheimer95} and the first hot Jupiter planets
beginning with 51 Peg b \citep{mayor95,marcy97}. Many more such objects have now
been discovered and reveal that planets and brown dwarfs populate an intermediate range of
temperatures not explored previously. This has led to the requirement to develop new methods to model
these atmospheres that cover the effective temperature range from below 1000K to more than
2000K. 

One widely used approach has been to adapt stellar atmosphere codes to handle the
lower temperatures encountered in exoplanets and brown dwarfs. Models of this
type are described for example by \citet{tsuji96}, \citet{allard01}, \citet*{barman01},
and \citet*{burrows03,burrows06}. An alternative approach, and the one we
follow in this paper, is to take models originally used for the atmospheres of
the Earth or other solar system planets and adapt them to handle the higher 
temperatures needed for
exoplanets and brown dwarfs. Such an approach has been described by \citet{marley02}
and \citet{fortney05} who use a model based on one originally used to model the
atmospheres of Titan and Uranus \citep{mckay89,marley99}.

One important difference between the techniques used in stellar atmosphere modelling and in
Earth atmosphere modelling is the approach to radiative transfer. In stellar atmospheres 
simplified treatments of scattering, such as the assumption of isotropic scattering, are
commonly adopted. This is justified by the fact that scattering is in most cases not an
important source of opacity in stellar atmospheres, and where it does become significant, in
the form of Rayleigh scattering from molecules in cool stars, and scattering from electrons
in hot stars, the phase functions are forward-backward symmetric. Scattering from clouds and
aerosols in the Earth atmosphere can, however, result in strongly forward peaked phase
functions, and to properly model such cases radiative transfer methods that more rigorously
handle multiple scattering with anisotropic phase functions are needed. While these techniques
build on the classic work of astronomers such as Chandrasekhar and van de Hulst, much recent
development of such methods has been in the context of Earth atmosphere research 
\citep[e.g.][]{liou02}.

Clouds are now known to be important not just in the atmospheres of all the solar system
planets, but in many brown dwarfs and even in late M dwarfs. Thus for all these objects a more
rigorous treatment of radiative transfer is desirable. Such an approach is particularly
important for modelling the reflected light from exoplanets, because the angular dependence
of scattering is an important factor in determining the phase variation around a planet's
orbit \citep[e.g.][]{seager00,cahoy10}. 

In this paper we describe the methods used in the VSTAR (Versatile Sofwtware for Transfer of Atmospheric
Radiaition) atmospheric modelling software. VSTAR was originally developed as a way of 
modelling the spectra of the solar system planets,
and an early version of it is described by \citet{bailey06}. In this paper we describe the
current version of VSTAR which can now handle a wide range of atmospheres ranging from those
of the coolest solar system planets up to stars with temperatures of $\sim$3000K, and thus
including the brown dwarfs and hot Jupiter type exoplanets.

\section{The VSTAR Model}

\subsection{Ionization and Chemical Equilibrium}
\label{sec_ice}

The Ionization and Chemical Equilibrium (ICE) package of VSTAR is used to
determine the equilibrium chemical composition of an atmospheric layer given its
elemental abundances, pressure and temperature. ICE handles gas phase
chemistry, ionized species and the formation of solid and liquid condensates.
Full details of the methods employed in this package will be described
elsewhere so only a brief description is given here. The techniques used are
similar to those described by \citet{tsuji73}, \citet{allard01} and
\citet{lodders02}. For each compound considered in the model the equilibrium
constant of formation $K_f$ from the elements is required. These are used in a 
set of equations for the mass balance of each element, and for the charge balance
to solve for the abundances of each molecular species. $K_f$ is related to
the Gibbs free energy of formation $\Delta_fG^o$ through:

\begin{equation}
\Delta_fG^o = -RT\ln{K_f}
\end{equation}

Where R is the gas constant.
This relationship is indicative of the link between the technique we use based on
equilibrium coinstants, and the alternative technique for chemical models based on
minimization of the total Gibbs free energy of the system \citep{sharp90,sharp07}
Both $\Delta_fG^o$ and $K_f$ are functions of temperature (T) and are available in
standdard compilations of thermochemical data. Our main source of thermochemical
data was the fourth edition of the NIST-JANAF thermochemical tables 
\citep{chase98}. We used data from \citet{lodders99} and \citet{lodders04} for
several compounds for which the JANAF data have been shown to contain errors.
Additional sources of thermochemical data were \citet{barin95} and
\citet{robie95}. 

Some important species, however, do not have data in any of these tabulations, and for
some others the available data do not extend to sufficiently high temperatures for
stellar atmosphere models. For a number of gas-phase species we have therefore
calculated our own thermochemical data. To do this we first calculate the partition
function using the spectroscopic constants of the molecule.  We use a rigid rotator
harmonic oscillator model (with additonal corrections for anharmonicity and
centrifugal distortion) for the rotational and vibrational levels, and direct
summation over the electronic levels \citep{mayer40}. The thermodynamic functions can
then be derived from the partition function and the dissociation energy of the
molecule. The techniques are essentially the same as those used in constructing the
NIST-JANAF thermochemical tables \citep{chase98} for gas phase species, and we have
tested our methods by reproducing results published in these tables. Species we have
calculated new thermochemical data for include FeH, CrH, CaH, TiH, RbCl and RbF.

\begin{figure}
\includegraphics[width=74mm,angle=270]{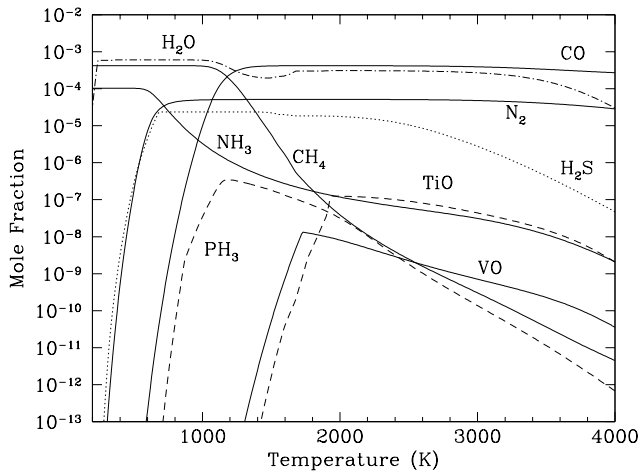}
\caption{Mixing ratios on several important species as a function of temperature for a
solar composition gas at 1 atm pressure calculated with our chemical equilibrium model. The
results can be compared with figure 17 of \citet{sharp07}}
\label{fig_abund}
\end{figure}

Our thermochemical model currently includes 143 gas-phase and condensed-phase
compounds  of 27 elements. While this is not as extensive as some other models, we
have been careful to include all species listed as important in previous studies such
as \citet{burrows99}. Our chemical model predictions show good agreement with previous
results such as those of \citet{lodders02} and \citet{sharp07}. Figure \ref{fig_abund}
shows the results of calculations of the mixing ratios, or mole fractions, for a
number of important species at 1 atmosphere pressure over a range of temperatures. It
can be directly compared with figure 17 of \citet{sharp07}. The model can easily be
expanded by adding new species to our thermochemical database.

Our current model assumes true equilibrium chemistry in that when condensates form, the condensed phase is
assumed to remain in equilibrium with the gas phase. This assumption is commonly made in such
models. However, it may not be true in real cases since condensed material can fall under gravity
and rain out of the system. Models that take account of this "rainout" process are described by
\citet{marley02} and \citet{freedman08}. 

\subsection{Molecular Absorption Lines}
\label{sec_mol}

Absorption lines due to rovibrational and electronic transitions of molecules are the
most important features of the spectra of planets, brown dwarfs and the coolest stars.
For use with VSTAR we have collected a line database that currently contains more than
2.9 billion lines. The  line lists used with VSTAR are listed in table
\ref{tab_linelists}. The line parameters are in a variety of different formats. The
essential data needed for all lines are the line position (as wavelength or
wavenumber), the line intensity, and the lower state energy of the transition
(ususally given in cm$^{-1}$ or eV).

Line intensity is usually given in one of three forms. HITRAN, GEISA (see section
\ref{sec_higeis}) and similar lists
give a line intensity $S_0$ in units of cm molecule$^{-1}$ (sometimes given as cm$^2$
mol$^{-1}$ cm$^{-1}$) at a reference temperature $T_0$. For HITRAN $T_0$ is 296 K, but
some other lists use a different reference temperature.

The line intensity at temperature $T$ is then calculated using \citep{rothman98}:

\begin{equation}
S = \frac{S_0 Q(T_0)}{Q(T)} \frac{\exp(-c_2 E_l/T)}{\exp(-c_2 E_l/T_0)} \frac {[1 - \exp(-c_2
\nu_0/T)]}{ [1 - \exp(-c_2\nu_0/T_0)] }
\label{eqn_hitran}
\end{equation}

where $\nu_0$ is the line frequency in cm$^{-1}$, $E_l$ is the lower state energy in
cm$^{-1}$, $c_2$ is the second radiation constant ($ = hc/k$) and $Q(T)$ is the
partition function (or total internal partition sum).

Line intensities may also be quoted in the form of Einstein A coefficients ($A_{21}$).
The line intensity S (in cm/mol) at a temperature T can be calculated from $A_{21}$
using \citep{simeckova06}.

\begin{equation}
S = \frac{g A_{21}}{8 \pi c \nu_0^2 Q (T)} \exp(-c_2 E_l/T) (1-\exp(-c_2 \nu_0/T))
\label{eqn_acoef}
\end{equation}

Here $g$ is the statistical weight of the upper level of the transition. In most cases
$g$ can be written as $(2J + 1) g_s$ where J is the rotational quantum number of the
upper state, and $g_s$ is the nuclear spin degeneracy. Different formulae are needed
if hyperfine structure is included in the line list \citep[see][]{simeckova06}. The
g$_s$ used in equation \ref{eqn_acoef} needs to be consistent with that used in
calculating the partition function (see section \ref{sec_part_nuc}).

In astronomical line lists, such as those of \cite{kurucz05}, the oscillator strength $f$, usually tabulated as $gf$ or
$\log(gf)$ (where $g$ is the statistical weight) is the usual form for listing line
strengths. The line intensity at temperature T can be calculated from $gf$ using:

\begin{equation}
S = \frac{\pi gf e^2}{m_e c^2 Q (T)} \exp(-c_2 E_l/T) (1-\exp(-c_2 \nu_0/T))
\label{eqn_gf}
\end{equation}

where $e$ is the electron charge, $m_e$ is the electron mass and other symbols are as
above.

\begin{table*}
\caption{Molecular Line Lists used with VSTAR}
\begin{tabular}{llrl} \hline
Molecule & List & Number of Lines & Reference \\\hline
39 molecules & HITRAN 2008 & 2,713,968   & \citet{rothman09} \\
50 molecules & GEISA 2009  & 3,807,997   & \citet{husson08} \\
H$_2$O       & BT2         & 505,806,202 & \citet{barber06} \\
H$_2$O       & SCAN        & 101,455,143 & \citet{jorgensen01} \\
H$_2$O       &             & 65,912,356  & \citet{partridge97} \\
H$_2$O       & HITEMP      & 111,377,777 & \citet{rothman10} \\
HDO          & VTT         & 697,454,528 & \citet{voronin10} \\
CO$_2$       &             & 7,088,178   & \citet{pollack93} \\
CO$_2$       & HITEMP      & 11,377,777  & \citet{rothman10} \\
CO$_2$       & CDSD-296    & 419,610     & \citet{tashkun03} \\
CO$_2$       & CDSD-1000   & 3,950,553   & \citet{tashkun03} \\
CO$_2$       & CDSD-Venus  & 11,730,277  & \citet{tashkun03} \\
CO           & HITEMP      & 115,218     & \citet{rothman10} \\
CO           &             & 134,421     & \citet{goorvitch94} \\
CH$_4$ (cool)&             & 339,690     & see text (section \ref{sec_ch4})\\
CH$_4$ (hot) &             & 134,862,336 & see text (section \ref{sec_ch4})\\
NH$_3$       &             & 3,249,988   & \citet{yurchenko09} \\
NH$_3$       & BYTe        & 1,138,323,351 & \citet*{yurchenko11} \\
TiO          & SCAN        & 12,837,150  & \citet{jorgensen94} \\
TiO          &             & 37,744,499  & \citet{schwenke98} \\
TiO          &             & 11,369,552  & \citet{plez98} \\
VO           &             & 3,171,552   & Plez, B., private communication \\
CaH          &             & 124,615     & see text (section \ref{sec_mh}) \\
MgH          &             & 23,315      & \citet{weck03}; \citet{skory03} \\
MgH          &             & 119,167     & \citet{kurucz05} \\
FeH (F-X)    &             & 116,300     & \citet{dulick03} \\
FeH (E-A)    &             &   6,357     & \citet{hargreaves10} \\
CrH          &             & 14,255      & \citet{burrows02} \\
TiH          &             & 199,073     & \citet{burrows05} \\
CH           & SCAN        & 114,567     & \citet{jorgensen96} \\
CN           & SCAN        & 2,245,378   & \citet{jorgensen90} \\
C$_2$        &             & 360,887     & \citet*{querci71}; \citet*{querci74} \\
HCN/HNC      &             & 34,433,190  & \citet{harris06} \\
\hline
\end{tabular}
\label{tab_linelists}
\end{table*}

\subsubsection{HITRAN and GEISA}
\label{sec_higeis}

HITRAN (HIgh resolution TRANsmission) and GEISA (Gestion et Etude des Informations
Spectroscopiques Atmosph\'{e}riques) are compilations of molecular spectroscopic
parameters designed primarily for the Earth atmosphere, but often useful for the
atmospheres of other planets. Both are updated every few years with the most recent
releases being HITRAN 2008 \citep{rothman09} and GEISA 2009 \citep{husson08}. The line
parameters are generally best suited to low temperature models and more complete line
lists are usually needed for higher temperature atmospheres such as those encountered
in brown dwarfs, hot Jupiter type planets and stars. For example \citet{bailey09}
showed that HITRAN and GEISA were significantly incomplete for water vapour line
parameters at temperatures of 500-700K encountered in the Venus lower atmosphere.

\subsubsection{HITEMP}

The HITEMP database is a companion to HITRAN containing line data suitable for use at higher
temperatures. The latest edition of HITEMP was released in 2010 \citep{rothman10} and replaces
an earlier edition described by \citet{rothman95}. HITEMP contains data for five species,
H$_2$O, CO$_2$, CO, NO and OH. The data format is the same as that used in HITRAN, and HITEMP
is consistent with HITRAN, in the sense that lines common to both databases have the same
line parameter values, although HITEMP includes many lines not in HITRAN. 

\subsubsection{Water Vapour --- H$_2$O}

Water vapour is an important absorber in atmospheres ranging from stars to the
terrestrial planets. A number of line lists are available for water vapour at high
temperatures. These include the HITEMP list \citep{rothman10}, the SCAN list
\citep{jorgensen01}, the list of \citet{partridge97} and the BT2 list \citep{barber06}. 
Comparisons of the various lists have been
reported by \citet{allard00}, \citet{jones03} and \citet{bailey09}. We normally use
the BT2 list which is the most extensive and is based on the acurate dipole
moment surface of \citet{schwenke00}. The HITEMP list for H$_2$O is based on BT2.

BT2 includes only the main isotopologue of water (H$_2$$^{16}$O). The VTT list of
\citet{voronin10} provides a similar list for HDO. In the atmosphere of Venus
deuterium is enhanced over terrestrial abundances by a factor of 100--150 so HDO
absorption is significant. Line data for a range of other isotopologues of water at 296K and
1000K are available from the Spectra Information System (http://spectra.iao.ru) at the
Institute of Atmospheric Optics, Tomsk. These data are based on the analysis of
\citet{schwenke00}.

\subsubsection{Carbon Dioxide --- CO$_2$}

Carbon dioxide is an important constituent of terrestrial planet atmospheres. Until
recently it has not been thought to be important in giant planet atmospheres. However,
recent results from transitting extrasolar planets \citep{swain09a,swain09b} suggest
CO$_2$ may be important in these  atmospheres as well. Recently CO$_2$ has also been
detected in brown dwarfs \citep{yamamura10}. At low temperatures HITRAN,
GEISA or CDSD-296 \citep[Carbon Dioxide Spectroscopic Databank, ][]{tashkun03} can be
used. High temperature line lists include the list described by \citet{pollack93} and
 the CDSD-1000 list \citep{tashkun03}.
The latter list contains more than 3 million lines and is complete to an intensity of
10$^{-27}$ cm mol$^{-1}$ at 1000 K.

The dense CO$_2$ atmosphere of Venus provides a particular challenge to the modelling
of CO$_2$ absorption with pressures up to 90 bars of almost pure CO$_2$ and requires a
line list with a much deeper intensity cutoff than is normally necessary. The list of \citet{pollack93} is the one
that has normally been used for modelling the Venus deep atmosphere. However, an
alternative is the CDSD-Venus list, a version of the CDSD list with a deeper cutoff of
10$^{-30}$ cm mol$^{-1}$ at 750 K.

The HITEMP \citep{rothman10} list for CO$_2$ is based on the CDSD line lists.

\subsubsection{Carbon Monoxide --- CO}

Carbon monoxide is significant in terrestrial planet atmospheres, and is also
observed in Titan. It also becomes
important in high temperature atmospheres where equilibrium chemistry favours CO over
CH$_4$. For high temperatures the HITEMP line list \citep{rothman10} or the list of
\citet{goorvitch94} can be used.

\subsubsection{Methane --- CH$_4$}
\label{sec_ch4}

Methane is an important absorber in the atmospheres of the solar system giant planets
and in Titan. It has been detected in the atmospheres of extrasolar giant panets
\citep{swain08}. It is also important in brown dwarfs, the presence of methane
absorption features in the near-IR being the defining characteristic of the T-dwarf class.

However, methane has a complex spectrum due to the presence of coincidences between
its four vibrational modes that result in a series of interacting states known as
polyads, spaced at intervals of about 1500 cm$^{-1}$. Most line data is either based
on experimental measurements or on effective Hamiltonian models, which exist only for the lowest few
polyads. The Spherical Top Data System (STDS) software of \citet{wenger98} can be used
to generate line lists from these effective Hamiltonian models. A good model has
recently been obtained \citep{albert09} for the ground-state and the lowest three polyads
(the Dyad, Pentad and Octad)
 and this allows reliable prediction of the low temperature methane
spectrum from 0 -- 4800 cm$^{-1}$. This model is used in the 2008 update of the
methane line parameters in HITRAN \citep{rothman09}. A preliminary model is available for
the next polyad, the Tetradecad \citep{boudon06,robert01}.

Above 4800 cm$^{-1}$ many of the line parameters included in HITRAN are empirical
measurements at room temperature originating from \citet{brown05}. These mostly lack
lower state energies (HITRAN lists fictitious lower state energies of 555.5555 or
333.3333) and so cannot be used to derive reliable line intensities at other
temperatures. However, some lines in the 5500 -- 6150 cm$^{-1}$ region have empirical
lower state energies derived from measurements at multiple temperatures from
\citet{margolis90} and \citet{gao09}.

Recently much improved data for the low temperature methane spectrum over the range 1.26 --
1.71 $\mu$m (5852 --- 7919 cm$^{-1}$ has become available from laboratory measurements described by
\citet{wang10,wang11}, \citet{campargue10} and \citet{mondelain11}. These data are based on a
combination of room temperature and cryogenic ($\sim$80K) measurements that allow empirical
determinations of lower state energies, and includes deep cavity ring down spectroscopy of the
weak lines in the window regions between the strong absorption band systems. An application of
this list to the spectrum of Titan is reported by \citet{debergh11}.

For low temperature atmospheres,
such as the solar system giant planets or Titan, we use a composite list that combines the
laboratory data described above, as well as data described by \citet{nikitin10, nikitin11},
with HITRAN data from 0 -- 4800 cm$^{-1}$. We use STDS to fill in the remaining gap in
coverage between 4800 and 5500 cm$^{-1}$. An early version of this list and its application to
Titan is described by \citet{bailey11}. The current version of the list is further improved by
including the most recent data of \citet{wang11} and \citet{mondelain11}.

For high temperature atmospheres, such as those of brown dwarfs and hot Jupiters, we
use a line  list computed with the STDS software \citep{wenger98} up to J=60 for most band systems, and
up to J=50 for the most complex system modelled (the Tetradecad-Pentad). This gives a
list of nearly 135 million lines. Because the effective Hamiltonian model of
\citet{albert09} involves high order polynomial fits to empirical data usually limited
to the range up to J=20-30, it is not suitable for extrapolation to these high J
values. We instead use the low order effective Hamiltonian parameters of
\citet{borysov02}. These provide a poorer fit to observed line positions ($\sim$ 0.5
cm$^{-1}$) but allow more reliable extrapolation to high J. For most purposes we use a
smaller list extracted from the full list with a line intensity cutoff of 10$^{-27}$
cm mol$^{-1}$ at 1500 K giving a list of 15.8 million lines.

While this approach to the hot methane spectrum is similar to that used by other
groups modelling brown dwarf and exoplanet spectra 
\citep[e.g.][]{freedman08,homeier03,borysov02} it is important to understand that
while these models include high rotational levels, they do not include higher
vibrational levels that are needed at these temperatures and many hot band systems are
therefore missing. Methane band strengths will therefore be underestimated by amounts
that will increase for shorter wavelengths and higher temperatures. 

Above 6500 cm$^{-1}$ there is little useful methane line data for high 
temperatures. The low temperature line list can be used but will omit both hot bands and
high J transitions. An alternative is
available in the form of low resolution absorption coefficients, band models or
k-distribution parameters for methane \citep{strong93,irwin06,karkoschka10}. The first
two sets cover the near-IR, and the latter extends through the visble as well.
However, these data are not in an ideal form for use with line-by-line programs such as
VSTAR and are not designed for high temperature use. The optimal way to use such datasets is in conjunction with the correlated-k
method \citep{goody89}.

\subsubsection{Ammonia --- NH$_3$}

Ammonia is present in the atmospheres of Jupiter and Saturn, and is also found in cool
brown dwarfs where it is seen most easily through a feature at 10.5 $\mu$m
\citep{cushing06}. The spectrum of ammonia, has until recently, presented
similar problems to methane. The
line parameters in HITRAN only include lines up to 5295 cm$^{-1}$ and have not been
updated since the 2000 edition. However, recently a new line list (the BYTe list) for hot NH$_3$ (up to $\sim$1500 K)
has been published by \citet{yurchenko11}. It contains more than 1.1 billion lines covering the
range 0 -- 12,000 cm$^{-1}$. An NH$_3$ list for use up to 300K and covering 0 -- 8000 cm$^{-1}$
 was described by \citet{yurchenko09}. 

\subsubsection{Metal Oxides --- TiO and VO}

Bands of Titanium Oxide (TiO) and Vanadium Oxide (VO) are distinctive features of the
spectrum of M-type stars. There are several line list available for TiO from
\citet{jorgensen94}, \citet{schwenke98} and \citet{plez98}. The lists cover five
isotopologues of the TiO molecule. Comparisons of the lists
can be found in \citet{allard00} and \citet{pavlenko06}. The VSTAR database also includes a
VO line list provided
by Plez, B. (private communication), which is part of the line lists assembled by
\citet{gustafsson08}.

\subsubsection{Metal Hydrides --- CaH, MgH, FeH, CrH and TiH}
\label{sec_mh}

Bands due to electronic transitions of metal hydrides are seen in M-type stars and some
cooler objects. A line list for MgH is described by \citet{weck03} and \citet{skory03}.
This list, and a similarly formatted list for CaH (which does not appear to have a
published description) can be found on the web site of the University of Georgia
Molecular Opacity Project\footnote{http://www.physast.uga.edu/ugamop/index.html}. An
alternative MgH list which includes two additional MgH isotopologues is available from
\citet{kurucz05}.

Lines lists are also available for FeH \citep{dulick03}, CrH \citep{burrows02} and TiH
\citep{burrows05}. These line lists are all similarly formatted. They cover only the
most abundant isotopologue of each molcule, though methods for determining line positions
for other isotoplogues are described. The \citet{dulick03} line list for FeH covers the F-X
band, but absorptions in the E-A band are also important in the 1.6 $\mu$m region. These are
available from a recent empirically based line list described by \citet{hargreaves10}.

\subsubsection{CH, CN and C$_2$, HCN/HNC}

These bands are present in cool stars and become particularly strong in carbon stars.
Line lists available from \citet{jorgensen96} (CH), \citet{jorgensen90} (CN),
\citet{querci71} and \citet{querci74} (C$_2$) and \citet{harris06} (HCN/HNC) are included in the
VSTAR database.

\subsubsection{Absorption Plots}

\begin{figure}
\includegraphics[width=90mm]{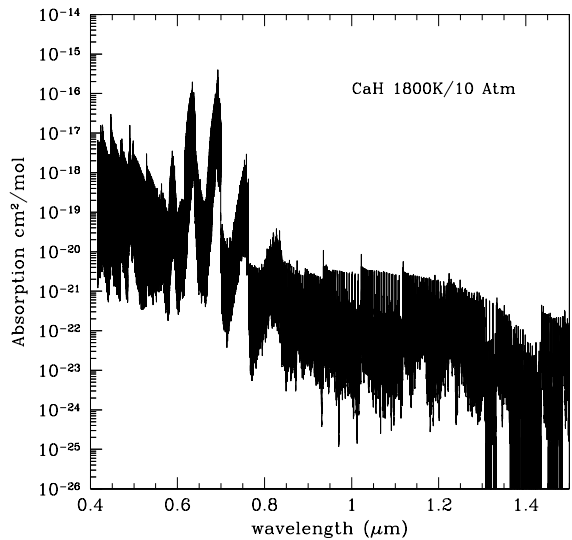}
\caption{Monochromatic absorption (in cm$^2$ molecule$^{-1}$) of CaH at a temperature
of 1800 K and a pressure of 10 atm}
\label{fig_cah}
\end{figure}

\citet{sharp07} have presented plots of the monochromatic absorption (in cm$^2$
molecule$^{-1}$) for many molecular species present in brown dwarf or giant exoplanet
atmospheres. We have made similar plots for the molecular species described above. In all but
one case our plots look essentially the same as those given by \citet{sharp07}. The exception
is CaH where our absorption plot, given in figure \ref{fig_cah}, shows
the well known absorption bands at 0.638 and 0.683 $\mu$m in M dwarfs, which are absent
in figure 9 of \citet{sharp07} for the same temperature and pressure. 

\subsection{Atomic Lines}
\label{sec_atom}

While molecular lines are most important in the coolest objects, lines of the alkali
metals are significant in brown dwarfs and exoplanets, and many other atomic lines
become important in M stars. VSTAR can make use of atomic line data in the format of
the Kurucz line lists, or that of the Vienna Atomic Line Database \citep[VALD,
][]{piskunov95,kupka99}.

\subsection{Line Profiles}

The default treatment of spectral line profiles in VSTAR is to use a Voigt line shape
in the line core, approximated using the methods described by \citet{humlicek82} and
\citet{schreier92}, out to 20 Doppler widths from the line centre. In the line wings a
van-Vleck Weisskopf profile is used which has a line shape function $\phi(\nu)$ of
the form \citep{vanvleck45}:

\begin{equation}
\phi(\nu) = \frac{1}{\pi} \left(\frac{\nu}{\nu_0}\right) \left[\frac{\gamma}{(\nu-\nu_0)^2+\gamma^2} +
\frac{\gamma}{(\nu+\nu_0)^2+\gamma^2}\right]
\label{eqn_vvw}
\end{equation}

Where $\nu$ is frequency, $\nu_0$ is the line centre frequency, and $\gamma$ is the
line half width. In most cases this is essentially identical to a Lorentzian, but this
profile provides the asymmetry that is needed at microwave wavelengths or with very
broad lines, and goes to zero at zero frequency.

In general it is found that continuing this profile out to large distances from the
line centre overestimates the absorption in the far wings. In cases where the precise
far wing behaviour is important, VSTAR allows the line wings to be modified by
multiplying by a correction factor, usually called a $\chi$ factor. For example the
far line wings of CO$_2$ are known to be sub-Lorentzian \citep{perrin89,tonkov96}, and
the correct profile is important in modelling the spectra of the ``windows'' in the
Venus nightside spectrum \citep[e.g.][]{meadows96}.

Where the precise line wing shape is not critical, it is usually sufficient to cut
off the line profiles at a distance from the line centre, which can be chosen
according to the typical line strengths and widths but usually ranges from 20 to
100 cm$^{-1}$. This approach improves the speed of computation. In either this case, 
or the $\chi$
factor case, the line intensities are adjusted to ensure that the integrated line
strength is not changed from its original value.

The line profile requires a value for the pressure broadened line half width
($\gamma$ in equation \ref{eqn_vvw}). HITRAN, GEISA and
similarly formatted files such as HITEMP and CDSD include line width data in the form of an
air broadened line half width coefficient $\gamma_0$ (in cm$^{-1}$ atm$^{-1}$, quoted for
the reference temperature $T_0$), a self
broadened line half width, and a temperature exponent $n$ for the line width such that:

\begin{equation}
\gamma (T) =  \gamma_0 \left(\frac{T_0}{T} \right) ^n
\end{equation}

Line widths for different transitions of a molecule vary, and are usually found to depend
primarily on the rotational quantum numbers of the levels involved in the transition. 
Variation with the vibrational quantum numbers is much smaller, and is often ignored in
empirical models for the line shape.

Line widths depend on the type of gas responsible for the broadening. Hence the air broadened
widths given in HITRAN type files are not usually what is needed, as we generally want the
line widths for broadening in H$_2$ and He (for giant planet and brown dwarf atmospheres)
or CO$_2$ (for terrestrial planets). Unfortunately the data available on these alternate
broadening gases is rather incomplete.

\begin{table*}
\caption{Relative broadening effects of different broadening gases}
\begin{tabular}{lllllllll}
Lines & \multicolumn{7}{c}{Broadened by} & References \\
of       & H$_2$ & He & N$_2$ & O$_2$ & CO$_2$ & Air & H$_2$/He &   \\ \hline
NH$_3$   & 0.832 & 0.370 & 1.0 & 0.693 &    &  0.939 & 0.818 &  \citet{brown94} \\
CH$_4$   & 1.017 & 0.643 & 1.0 & 0.943 &    &  0.989 & 0.961 &  \citet{pine03} \\ 
CO       & 0.85  & 0.64  & 1.0 & 0.90  & 1.30 &  0.98  & 0.82  &  \citet{burch62},
\citet{hartmann88} \\
H$_2$O   & 0.777 & 0.221 & 1.0 & 0.661 & 1.50 & 0.932 & 0.694 & \citet{gamache96} \\
CO$_2$   & 1.41  & 0.59  & 1.0 & 0.81  & 1.20 & 0.962 & 1.287 & \citet{burch69} \\
\hline
\end{tabular}
\label{tab_broad}
\end{table*}

Table \ref{tab_broad} shows the relative line widths due to broadening in different
broadening gases, averaged over many transitions. The line widths are relative to that due
to broadening in N$_2$.  Some general trends can be identified in the table.
For example, helium always produces the smallest broadening. O$_2$ broadening is always less than that
due to N$_2$. CO$_2$ produces the largest broadening, However, H$_2$ is sometimes less than, and
sometimes more than N$_2$. These trends can be used as a rough guide,
when no broadening information is available for a specific broadening gas. 

Broadening for a mixture of gases can be obtained by averaging the broadening coefficients
weighted by their partial pressures (or volume mixing ratios). Table \ref{tab_broad} gives
values for air (taken as 80\% N$_2$, 20\% O$_2$) and an H$_2$/He mixture typical of a
giant planet or brown dwarf atmosphere (85\% H$_2$, 15\% He) obtained from the other values
in the table.	 

VSTAR includes specific models for the line widths of several of the more important gases. For
H$_2$O broadened in CO$_2$ we use the model described by \citet{bailey09} based on the data
of \citet{delaye89}. For H$_2$O broadened in H$_2$ and He we use the data from
\citet{gamache96}. For NH$_3$ broadened by H$_2$, He, N$_2$ and O$_2$ we use the results of
\citet{brown94} for the dependence of width on the J and K quantum numbers, and the 
temperature exponents are from \citet{nouri04}. For CO we use the model for the J
dependence from HITRAN \citep{rothman05} for air, and from \citet{regalia05} for H$_2$. 
When CH$_4$ line data are taken
from HITRAN, the air broadened width from HITRAN is used for H$_2$/He broadening also,
since table \ref{tab_broad} shows little difference in the two cases. For other sources of
CH$_4$ line data the broadening is based on data from \citet{pine92}.

Atomic line lists usually give width data in the form of three coefficients for natural
($\gamma_n$), van der Waals  ($\gamma_w$) and Stark ($\gamma_s$) broadening. Stark
broadening is usually negligible at the temperatures considered here. The line width
$\gamma$ needed for equation \ref{eqn_vvw} can be calculated from:

\begin{equation}
\gamma = \frac{1}{4 \pi c}\left[\gamma_w(N_H + c_{1} N_{H_2} + c_{2}
N_{\mbox{\scriptsize He}})
\left(\frac{T}{10,000}\right)^{0.3} + \gamma_n\right]
\end{equation}

where $N_H$, $N_{H_2}$ and $N_{\mbox{\scriptsize He}}$ are the number densities of atomic hydrogen, molecular
hydrogen and helium. $c_{1}$ and $c_{2}$ are coefficients for the relative
broadening effects of H$_2$ and He as compared with H, which can be derived from the
polarizabilities of the different species. If $\gamma_w$ and $\gamma_n$ are not available
they are calculated using methods described by \citet{kurucz81}.

\subsection{Partition Functions}
\label{sec_part}

The calculation of line intensities from data tabulated in the line lists using equations
\ref{eqn_hitran}, \ref{eqn_acoef} or \ref{eqn_gf} requires the partition function $Q(T)$ of the
molecule or atom defined by.

\begin{equation}
Q(T) = \sum g_i \exp\left( \frac{-E_i}{kT}\right)
\label{eqn_partfn}
\end{equation}

where $E_i$ is the energy of level i relative to the ground state, $g_i$ is its statistical weight,
and the sum is to be taken over all rotational, vibrational and electronic levels. 

\subsubsection{Molecular Partition Functions}

VSTAR uses molecular partition functions from a range of sources. Polynomial approximations for the
partition functions of many molecules for temperatures from 1000 K up to 9000 K or higher 
are given by \citet{sauval84} and \citet{irwin81}. A subroutine
for calculating the partition functions of all molecules and isotopologues in  HITRAN is provided
with that database and is described by \citet{fischer03}. The HITRAN partition functions are valid over
the temperature range from 70 to 3000 K. An updated version of the HITRAN partition functions
has recently been published by \citet{laraia11}, but the work described in this paper used the
earlier versions from \citet{fischer03}. Partition functions can be reconstructed from data in the
NIST-JANAF thermochemical tables \citep{chase98}. These tables do not list the partition functions
explicitly, but the thermochemical properties listed for gas phase substances are derived
from partition functions. The partition function can be calculated from the tabulated data using
methods described by \citet{irwin88}.

Partition functions can also be calculated directly from spectroscopic constants using
methods already described in section \ref{sec_ice}. In some cases it is possible to
calculate molecular partition functions by direct summation over energy levels as in the partition
functions for H$_2$O from \citet{vidler00} and CH$_4$ from \citet{wenger08}.

\subsubsection{Partition Functions and Nuclear Spin}
\label{sec_part_nuc}

In comparing the molecular partition functions from different sources it was noticed that the
partition functions from HITRAN \citep{fischer03} had values which were larger than those from most
other sources by integer factors. This is due to the use in HITRAN of a different convention for
the treatment of
nuclear spin states. The convention in astrophysics is not to count nuclear spin states as distinct
states in equation \ref{eqn_partfn}. As stated by \citet{irwin81}, ``the statistical weights are
divided by ... the product of the nuclear spin statistical weights''. It is clear from the
descriptions of the HITRAN partition functions in
\citet{fischer03}, \citet{simeckova06} and \citet{laraia11} that a different convention is adopted in HITRAN, and some
other spectroscopy literature, where the corresponding statistical weights include nuclear spin
states.

The consequence is that on the HITRAN convention, partition functions $Q_{\mbox{\scriptsize HIT}}$ are
numerically larger than astrophysical partition functions $Q_{\mbox{\scriptsize ast}}$ by a factor
defined as follows:

\begin{equation}
Q_{\mbox{\scriptsize HIT}} (T) = Q_{\mbox{\scriptsize ast}} (T) \prod_{j=1}^n (2I_j + 1)
\end{equation}

where $I_j$ are the nuclear spins of the $n$ atoms contained in a molecule. With $I$ = 0 for carbon
and oxygen,
$^1/_2$ for hydrogen and 1 for nitrogen, these factors becone 4 for H$_2$O, 16 for CH$_4$ and 24
for NH$_3$.

In many cases (e.g. equation \ref{eqn_hitran}) partition functions are divided such that these integer
factors will cancel out. However, this is not the case when line intensities are calculated from
Einsten A coefficients using equation \ref{eqn_acoef}. In this case it is important that the
statistical weights ($g_s$) and partition function use the same convention. For example for H$_2$O
lines, if an astrophysical partition function is used the statistical weights for the ortho and para
states of water are $^3/_4$ and $^1/_4$, but if the HITRAN partition function is used they are 3
and 1.

\subsubsection{Atomic Partition Functions}

Partition functions for atoms and atomic ions in VSTAR are calculated using a modified version
of the PFSAHA subroutine taken from the ATLAS9 software of Kurucz\footnote{http://kurucz.harvard.edu/programs/atlas9}.

\subsection{Continuum Processses}
\label{sec_cont}

A number of continuum absorption processes are included in VSTAR as described in table \ref{tab_cont}.

\subsubsection{Bound-Free and Free-Free Continuum Absorption}

Bound-free and free-free absorptions of a number of species are important in stellar atmospheres, and
some of these processes remain significant at the cooler temperatures of hot Jupiters and brown
dwarfs. The processes included in VSTAR are listed in table \ref{tab_cont}. These processes are sufficient
for the range of temperatures and wavelengths currently studied using VSTAR. A wider range of
continumm processes would need to be incorporated to extend the applicability of VSTAR to higher
temperatures and into UV wavelengths \citep[see e.g.][]{sharp07,gustafsson08}.

\subsubsection{Collision Induced Absorptions}
\label{sec_cia}

Collision induced absorption (CIA) due to H$_2$ - H$_2$ pairs is important in the solar system giant planets
where it shows up a distinct strong spectral feature at 2.1 $\mu$m (see figure \ref{fig_ss}
for the spectrum of Jupiter in this region). This absorption is also important
in brown dwarf, hot jupiters, and M dwarfs. We incorporate this absorption by interpolating in tables
provide by Borysow\footnote{http://www.astro.ku.dk/~aborysow/programs}. Three different tables are
needed to cover the full temperature range using calculations by \citet{borysow02} (60 -- 350 K and 400 -- 1000 K) and
\citet{borysow01} (1000 -- 7000 K). H$_2$ - He collision induced absorption is also included using
data from \citet{borysow97}, \citet{borysow89} and \citet{borysowf89}.

Other collision induced absorptions included in VSTAR are absorption in the O$_2$ near infrared bands
\citep{smith99,smith00} which is needed for the Earth atmosphere, and H$_2$ - N$_2$ collision induced
absorption which occurs in the Titan atmosphere and for which data is available from
\citet{mckellar89}.

\begin{table}
\caption{Continuum absorption processes in VSTAR} 
\begin{tabular}{ll}
Process & Reference \\ \hline
H$^-$ free-free & \citet{bell87} \\
H$^-$ bound-free & \citet{wishart79} \\
H bound-free and free-free & \citet{gray05} pp149-154 \\
H$_2^-$ free-free & \citet{bell80} \\
H$_2$ - H$_2$ CIA & \citet{borysow02} \\
                  & \citet{borysow01} \\
H$_2$ - He CIA & \citet{borysow97,borysow89} \\
                   & \citet{borysowf89} \\
O$_2$ - O$_2$ and O$_2$ - N$_2$ CIA & \citet{smith99,smith00} \\
H$_2$ - N$_2$ CIA & \citet{mckellar89} \\
\hline
\end{tabular}
\label{tab_cont}
\end{table}
. 

\subsection{Scattering}
\label{sec_scatt}

Scattering processes are often treated in stellar atmosphere codes as just an additional source
of opacity which can be added to that due to gas absorption. However, VSTAR uses a more rigorous
treatment that requires a full description of scattering. In addition to the
optical depth contribution, we need to know the single scattering albedo, $\varpi$ and the 
phase function, $P(\theta)$, that describes the angular distribution of scattered  light.
  
Single scattering albedo is defined as: 

\begin{equation}
\varpi = \frac{\sigma_s}{\sigma_s+\sigma_a}=  \frac{\sigma_s}{\sigma_e}
\label{eqn_ssa}
\end{equation}

where 
$\sigma_e$,  $\sigma_a$ and $\sigma_s$ are the extinction, absorption and scattering cross 
sections per particle.

\subsubsection{Rayleigh Scattering from Molecules}
\label{sec_ray}

VSTAR includes models for the Rayleigh scattering due to air, H, H$_2$, He, N$_2$ and CO$_2$ or any
mixture of these gases. The wavelength dependence of the scattering cross section of H$_2$ is from
\citet{dalgarno62}. Those of H and He are from Dalgarno as cited by \citet{kurucz70}.

For the other gases the scattering cross section ($\sigma$) is derived from the refractive index of the gas
using \citep{hansen74}:

\begin{equation}
\sigma = \frac{8 \pi^3 (n(\lambda)^2 - 1)^2}{3 \lambda^4 N^2} F_K 
\end{equation}

\noindent where $n(\lambda)$ is the wavelength dependent refractive index, $\lambda$ is wavelength and N is the number density (molecules
cm$^{-3}$) of the gas measured at the same temperature and pressure as the 
refractive index. $F_K$ is known as the King factor and is given by:

\begin{equation}
F_K = \frac{6 + 3\delta}{6 - 7\delta}
\end{equation}

\noindent where $\delta$ is the depolarization factor. For air we use the refractive index wavelength dependence from
\citet{peck72} and the King factor from \citet{young81}. For CO$_2$ the refractive index is from
\citet{old71} and the King factor is from \citet{sneep05}. For N$_2$ the refractive index is from
\citet{cox00}.

Rayleigh scattering is pure scattering so it has a single scattering albedo of one, and the
phase function is:

\begin{equation}
P(\theta) = \frac{3}{4}(1 + \cos^2\theta)
\end{equation}

\subsubsection{Scattering from Particles}
\label{sec_spart}

Scattering from liquid or solid particles suspended in the atmosphere (variously called
clouds, aerosols or condensates) occurs in all planetary atmospheres in the solar system, and
in brown dwarfs and M stars.

Particle scattering can be modelled using Lorenz-Mie theory, which provides a
solution to Maxwell's equations that describes scattering of light from
a homogenous sphere with a complex refractive index $n + ik$, and a size parameter
$x = 2\pi r/\lambda$ where r is the radius of the 
sphere, and $\lambda$ is the wavelength. VSTAR uses a Lorenz-Mie scattering code
from \citet{mishchenko02} that models scattering from a size distribution of spherical
particles. The particle size distribution can be described by a number of functional
forms including power law, log normal and gamma distrbutions. VSTAR can handle up to 20
particle modes, each with its own wavelength dependent refractive index, particle size
distribution, and vertical distribution in the atmosphere.

The scattering code generates the extinction and scattering cross sections per particle
($\sigma_e$ and $\sigma_s$) which give the single scattering albedo according to equation
\ref{eqn_ssa}.

It also provides the phase function $P(\theta)$ both as an array of values for different
scattering angles, and as an expansion in Legendre polynomials

\begin{equation}
P(\theta) = \sum_{s=0}^{s_{\mbox{\tiny max}}}{ \alpha^s P_s(\cos{\theta})}
\label{eqn_mom}
\end{equation}

\noindent which is the form required for the discrete-ordinate radiative transfer code described in
section \ref{sec_rt}.

If the particles are genuinally spherical (e.g. liquid droplets) then the phase function
derived from the Lorenz-Mie code is appropriate. Sometimes however, spherical models are used simply
as a generic representation of solid particles which probably have a range of
irregular shapes In this case the full spherical particle phase function is probably not a good representation.
It will include angular structure in the form of features such as rainbows and glories, that
are specific to the particles' sphericity, and disappear for non-spherical particles
\citep{bailey07}. In this case a better representation may be the Henyey-Greenstein phase
function \citep{henyey41}:

\begin{equation}
P_{\mbox{\scriptsize HG}}(\theta) = \frac{1-g^2}{4 \pi (1+g^2-2g\cos{\theta})^{3/2}}
\end{equation}

\noindent where $g$ is the asymmetry parameter defined by:

\begin{equation}
g \equiv \langle\cos{\theta}\rangle = \int_0^{\pi} \! \cos{\theta} P(\theta) 2 \pi \sin{\theta}
\, d\theta
\end{equation}

Here $g$ is between $-$1 and 1, and is positive for forward scattering and negative for
backward scattering.
The asymmetry parameter is also provided by the Lorenz-Mie scattering
code.

As well as using Lorenz-Mie theory, VSTAR can make use of precalculated scattering
parameters derived using other methods. 
Scattering from non-spherical particles can be modelled using T-matrix methods
\citep{waterman71,mishchenko91}. Codes available from \citet{mishchenko02} can be used to
calculate the scattering properties of particles with spheroidal or cylindrical shapes 
or shapes described by Chebyshev polynomials in either random
or specific orientations. However, T-matrix
techniques are generally limited to small size parameters and can be very slow to compute.
Some examples of applications of these techniques can be found in \citet{bailey07} and
\citet{bailey08a}.

The aerosols that make up Titan's stratospheric haze have optical properties that can be
modelled by fractal aggregrates composed of many spherical particles. Models described by
\citet{tomasko08} give the scattering properties of the Titan aerosols including optical
depth, single scattering albedo and phase function in a form suitable for use with VSTAR.

\subsection{Radiative Transfer}
\label{sec_rt}

The data on absorption and scattering derived by the methods described in sections \ref{sec_mol}
to \ref{sec_scatt} are finally combined to provide the inputs needed for solving the radiative 
transfer equation. VSTAR is structured in a modular way so that a number of different approaches 
to radiative transfer can potentially be used. However, for most of our work we have used the DISORT package 
\citep{stamnes88}. DISORT is a robust computer implementation of the discrete ordinate 
method for radiative transfer originally developed by \citet{chandra60}.

DISORT models the radiative transfer in a multiple layer medium, including the processes of
absorption, multiple scattering and thermal emission, and allows a  reflecting surface at the lower boundary
and a direct beam source (e.g. the Sun) illuminating the top of the atmosphere. 

The radiative transfer equation solved by DISORT has the form:

\begin{equation}
\mu \frac{dI_{\nu}(\tau,\mu,\phi)}{d \tau} = I_{\nu} (\tau, \mu, \phi) - S_{\nu}
(\tau, \mu, \phi)
\end{equation}

\noindent where $I_{\nu}$ is the monochromatic radiance (sometimes referred to as intensity or specific
intensity) at frequency $\nu$, and is a function
of optical depth $\tau$, and direction $\mu$, $\phi$, where $\mu$ is the cosine of
the zenith angle, and $\phi$ is the azimuthal angle. The source function $S_{\nu}$ is
given by:

\begin{align}
S_{\nu} (\tau,\mu,\phi) & = \frac{\varpi(\tau)}{4 \pi} \int_0^{2 \pi} \!
\int_{-1}^{1} \! P(\mu,\phi; \mu', \phi') I_{\nu} (\tau, \mu', \phi') d \mu' d \phi' 
\nonumber \\
  &  \mbox{} + (1 - \varpi) B_{\nu} (T)   \\
  &  \mbox{} + \frac{\varpi F_{\nu}}{4 \pi} P(\mu,\phi; \mu_0, \phi_0) \exp{(-\tau / \mu_0)}
  \nonumber
\end{align}

\noindent where the first term describes scattering of radiation into the beam from other
directions according to single scattering albedo $\varpi$ and phase function $P(\mu,\phi; \mu', \phi')$, the second term
is thermal emission, with $B_{\nu} (T)$ being the Plank function and the third term is direct illumination of the atmosphere by
an external source with flux $\mu_0 F_{\nu}$ and direction $\mu_0, \phi_0$ (e.g. the Sun).   

The inputs required by DISORT
are the vertical optical depth ($\Delta\tau$), single scattering albedo ($\varpi$) and phase 
function moments ($\alpha^s$, the coefficients of the Legendre polynomial expansion of the phase
function as defined in equation \ref{eqn_mom}) for each atmospheric layer. The
temperature at each level is also needed for thermal emission calculations.

The values of these are obtained by summing the contributions of all relevant processes for each
layer of the atmosphere at each wavelength. The combined absorption due to all molecular and atomic
line processes as well as related continuum processes (sections \ref{sec_mol} to \ref{sec_cont}) is
called $\Delta\tau_{gas}$. Rayleigh scattering (section \ref{sec_ray}) contributes an optical depth
$\Delta\tau_{ray}$ which is pure scattering optical depth. Each of the $p$ modes ($p = 1$ to
$p_{max}$) of scattering particles
(section \ref{sec_spart}) provides both a scattering optical depth $\Delta\tau^p_{scatt}$ and an
absorption optical depth $\Delta\tau^p_{abs}$.

The total optical depth for a layer is:

\begin{equation}
\Delta\tau = \Delta\tau_{gas} + \Delta\tau_{ray} + \sum_{p=1}^{p_{\mbox{\tiny
max}}}{(\Delta\tau^p_{scatt} + \Delta\tau^p_{abs})}
\end{equation}

The combined single scattering albedo for a layer is the scattering optical depth divided by
the total optical depth so is given by:

\begin{equation}
\varpi = \frac{\Delta\tau_{ray} +
\sum_{p=1}^{p_{\mbox{\tiny max}}}{\Delta\tau^p_{scatt}}}{\Delta\tau}
\end{equation}

\noindent and the combined phase function moments ($\alpha^s$ for $s = 0$ to $s_{max}$) are the phase 
function moments for the individual scattering
components, the Rayleigh scattering ($\alpha^s_{ray}$) and the $p$ particle modes ($\alpha^{sp}$),
weighted according to their
contribution to the scattering optical depth.

\begin{equation}
\alpha^s = \frac{\alpha^s_{ray} \Delta\tau_{ray} +
\sum_{p=1}^{p_{\mbox{\tiny max}}}{\alpha^{sp}\Delta\tau^p_{scatt}}}{\Delta\tau_{ray} +
\sum_{p=1}^{p_{\mbox{\tiny max}}}{\Delta\tau^p_{scatt}}}
\end{equation}

DISORT approximates the angular distribution of the radiation field, by replacing the integral term
in the source function with a sum over a number of discrete zenith angles (or streams) according to the
Gaussian quadrature rule. The number of streams determines how accurately the angular distribution
of radiance is calculated. For many purposes where the variation with angle is reasonably smooth, we
have found 8 streams (four upward and four downward) to be adequate. However, the number of streams
can be increased as necesssary to provide more accurate representation of angular structure, at the
cost of increased computation time.

DISORT provides a number of outputs. In astronomical cases we are mostly interested in the upward
radiance and its angular dependence at the top of the atmosphere. The radiance can be determined for
any emission angle, and for any azimuth relative to the illuminating source. It is also possible to
determine the radiation field at the surface of the planet or at intermediate levels in the
atmosphere.

The radiative transfer equation only holds for monochromatic radiance. Thus to obtain the spectrum
of a planet the radiative transfer calculation must be repeated for each wavelength, and the
wavelength steps made sufficiently small to resolve the spectral lines. We normally use a set of
points that are equally spaced in wavenumber, and with a spacing chosen to be just sufficient to
resolve the narrowest absorption lines encountered in any layer of the model. Typically this requires
several hundred thousand spectral points to be calculated.

\subsection{VSTAR Operation}

We can now outline the full procedure for using VSTAR to calculate the spectrum of a planet, brown
dwarf or star. The first steps in the process are different depending on whether we are modelling
a solar-system planet (including the Earth) or an exoplanet, brown dwarf or star.

\subsubsection{First Step - Solar System case}

In the solar-system planet case we start with a measured profile for the temperature and pressure as
a function of height. This is available for all planets in the solar system, either from radio
occultation observations, or from entry probes or other in-situ data. For Earth there are a number
of ``standard atmosphere'' profiles available and several of these are built-in to VSTAR. Some other
planets have standard atmospheres, e.g. the Venus International Reference Atmosphere
\citep[VIRA,][]{seiff85}

The mixing ratios of the various gases and their distribution with altitude are also generally 
reasonably well known from in-situ measurement or past spectroscopic analysis. Thus for any such
planet we can start with an atmospheric profile which consists of a number of layers (typically
30 to 60) with the pressure, temperature, and mixing ratios of absorbing gases specified for each
layer. If clouds or aerosols are included in the model their vertical distribution is specified
in terms of the aerosol optical depth at a specific reference wavelength, for each particle mode
at each layer.

\subsubsection{First Step - Giant Exoplanet, Brown Dwarf or Star}

In these cases we don't have the direct measurements of profile and composition available for
solar system planets. Once again we start from a pressure-temperature profile for the atmosphere.
The other input required for the model is the elemental composition of the atmosphere. For this
we normally assume either a solar composition \citep[e.g.][]{grevesse07, asplund09, lodders09} or
a modified solar composition for a different metallicity.

Then for each layer of the atmosphere we use the ICE chemical equilibrium model (section
\ref{sec_ice}) to determine the equilibrium composition of the layer in terms of molecules, atoms
and ions. ICE also predicts what condensates are produced at different levels in the atmosphere,
and this can be used as a guide to the addition of clouds to the model which are specified in the
same way as in the solar system case.

Note that VSTAR is not an atmospheric structure model, and so cannot be used to determine
structures that are self consistent with the radiative transfer (although this capability may be
added in the future). Currently, therefore, we normally work from a pressure-temperature structure determined from
another stellar or exoplanet atmosphere model and VSTAR is used as a spectral synthesis model.

\subsubsection{Second Step - Computing Layer Absorption and Scattering Properties}

In either case we now have a specification of the atmospheric layers with their pressures,
temperatures, chemical composition and aerosol optical depth. We now calculate for each layer at
each modelled wavelength (typically there are several hundred thousand wavelength points) the
absorption and scattering properties. For line absorption this involves calculating the line
profiles of all the spectral lines of all relevant species that fall within the wavelength range
of the model (including lines outside the range where the far wing contribution may be
significant) and adding their contributions into the gas optical depth of the layer.

For particle scattering (clouds and aerosols - section \ref{sec_spart}) there are two options. Lorenz-Mie theory can
be used directly to calculate the scattering properties of each particle mode at each layer for
a set of wavelengths. These values are then spline interpolated to provide data for each
individual wavelength point. Alternatively precalculated scattering properties can be used and
interpolated in the same way.

\subsubsection{Third Step - Radiative Transfer Solution}

The next step is to combine the absoprtion and scattering properties of each layer as described
in section \ref{sec_rt} to provide the inputs needed for DISORT, and perform the radiative
transfer solution for each wavelength point. At this stage we specify the boundary conditions,
which can include a reflecting surface at the base of the atmosphere, and an illuminating source
(the Sun or another star) at a specified zenith angle.

VSTAR can calculate a number of different types of spectra. In the case of a solar system planet we
are ususally interested in the spectrum of the radiance at a specific point on the planet's disk, or
the radiance factor (I/F) if we are looking at reflected solar light. For a star we generally want
the flux spectrum (the radiance integrated over all angles) which is equivalent to what would be
seen from an unresolved star at a large distance. However, the angular dependence of the radiance
also allows the investigation of limb darkening of a star. VSTAR can also calculate the transmission
spectrum from the top of the atmosphere to the surface. This is used in particular for the Earth
atmosphere case where such data can be used to model telluric corrections to astronomical spectra
\citep{bailey07a}.
 
\section{Tests of VSTAR - Comparison with Other Models}

We have tested VSTAR by comparing its predictions with a number of other models.

\subsection{Comparison with RFM (Reference Forward Model)}

The Reference Forward Model (RFM)\footnote{http://www.atm.ox.ac.uk/RFM/} is a line-by-line model for the Earth atmosphere developed at
Oxford University for the MIPAS instrument project on the Envisat satellite. RFM is itself a development of the
earlier GENLN2 model \citep{edwards92}. RFM has been included in a number of intercomparison studies
of Earth atmosphere radiative transfer codes \citep{tjemkes03,saunders07}.

\begin{figure}
\includegraphics[width=90mm]{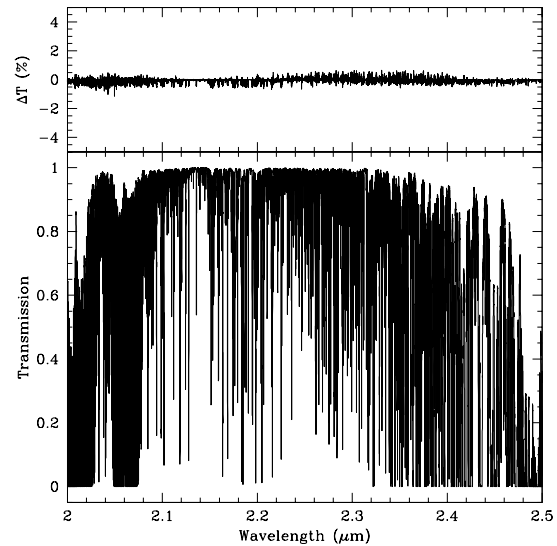}
\caption{Comparison of Earth atmosphere zenith transmission calculated with VSTAR and RFM in
the 2-2.5 $\mu$m window. Lower panel --- Transmission calculated with VSTAR. Upper panel ---
Percentage difference (VSTAR $-$ RFM).}
\label{fig_rfm_comp}
\end{figure}

We compared VSTAR with RFM (version 4.25) by using both codes to calculate the Earth atmosphere transmission
spectrum at the zenith in the 2 to 2.5 $\mu$m wavelength region. The atmospheric profile was the standard mid-latitude
summer atmosphere taken from the RFM website. Absorption due to H$_2$O, CO$_2$, O$_3$, N$_2$O, CO,
CH$_4$ and O$_2$ were included in the models with line data taken from the HITRAN 2000 database (the
version of HITRAN supported by RFM).

The transmission spectrum from VSTAR and the percentage difference between transmission calculated
by RFM and VSTAR are shown in figure \ref{fig_rfm_comp}. The mean difference (VSTAR $-$ RFM) in
transmission is $-$0.05\%, and the RMS difference is 0.09\%. There is no significant trend in the
differences with wavelength. These results are basically a test of the line-by-line absorption
calculations and show the two models are in good agreement. Residual differences between the two
codes appear to be due to differences in the line shape model used, and to differences in the way
layer properties are interpolated from the levels listed in the atmospheric profile.

\subsection{Radiative Transfer Benchmarks}

We have used VSTAR to reproduce a number of benchmark problems in radiative transfer
described by \citet{garcia85}. The first problem considered is a uniform Mie scattering atmosphere
containing spherical particles with size parameter 2 and index of refraction 1.33. The
atmosphere has a total optical depth $\tau$ = 1, and the single scattering albedo $\varpi$ =
0.95. The atmosphere is illuminated by a source at a zenith angle of 60 degrees and has a non-reflecting
surface at its base. The problem was implemented using VSTAR's own Mie scattering code to
calculate the phase function, and the results for the top of atmosphere upward radiances and
bottom of atmosphere downward radiances at an azimuth of zero are given in table
\ref{tab_bench}, and compared with the results summed from the fourier series coefficients
given in \citet{garcia85}. The VSTAR results, calculated with 16 streams in DISORT, agree
with the \citet{garcia85} results to five significant figures in all cases, and at over half the
zenith angles agree to within $\pm$1 in the sixth figure. Similar agreement is found for results at
other azimuths and optical depths.

Another problem we have calculated with VSTAR is one of the test problems posed by the 
Radiation Commission of the International Association  of Meteorology and Atmospheric 
Physics \citep{lenoble77}. This involves an atmosphere with $\tau$ = 1, $\varpi$ = 0.9
with a ``Haze L'' size distribution of scattering particles as defined by
\citet{lenoble77}. The atmosphere is illuminated by a source at a zenith angle of 60 degrees.
This problem has a more complex phase function, and with 16 DISORT streams the benchmark 
results in \citet{garcia85} are only matched to an accuracy of $\sim$1 per cent. However,
increasing the number of DISORT streams to 64 produces results that agree with the
\citet{garcia85} results to $\pm$1 in the sixth significant figure for the upward radiances
at the top of the atmosphere.

\begin{table}
\caption{Radiances for Mie scattering benchmark problem compared with values from
\citet{garcia85}. Positive values of $\mu$ are upward
radiances at the top of the atmosphere and negative values are downward radiances at the bottom of
the atmosphere, both for an azimuth of zero relative to the illuminating source.} 
\label{tab_bench}
\begin{tabular}{lllr}
\hline
$\mu = cos{zd}$ & VSTAR  &  Garcia \& Siewert & Difference \\
\hline
1.0    &   0.0476802   &   0.0476807   &   $-$0.0000005   \\
0.9    &   0.1072616   &   0.1072618   &   $-$0.0000002   \\
0.8    &   0.162275    &   0.162274    &      0.000001   \\
0.7    &   0.228132    &   0.228131    &      0.000001   \\
0.6    &   0.308466    &   0.308464    &      0.000002   \\
0.5    &   0.406536    &   0.406534    &      0.000002   \\
0.4    &   0.525328    &   0.525326    &      0.000002   \\
0.3    &   0.666624    &   0.666621    &      0.000003    \\
0.2    &   0.828749    &   0.828746    &      0.000003   \\
0.1    &   1.004042    &   1.004041    &      0.000001   \\
$-$0.1       &   0.466472    &   0.466478    &   $-$0.000006   \\
$-$0.2       &   0.578557    &   0.578561    &   $-$0.000004   \\
$-$0.3       &   0.653528    &   0.653530    &   $-$0.000002   \\
$-$0.4       &   0.682601    &   0.682601    &      0.000000   \\
$-$0.5       &   0.674533    &   0.674533    &      0.000000   \\
$-$0.6       &   0.637904    &   0.637903    &      0.000001   \\
$-$0.7       &   0.578009    &   0.578008    &      0.000001   \\
$-$0.8       &   0.496937    &   0.496936    &      0.000001   \\
$-$0.9       &   0.391880    &   0.391879    &      0.000001   \\
$-$1.0       &   0.197933    &   0.197932    &      0.000001   \\
\hline
\end{tabular}
\end{table}

\subsection{Comparison with MARCS}
\label{sec_marcs}

\begin{figure*}
\includegraphics[width=130mm]{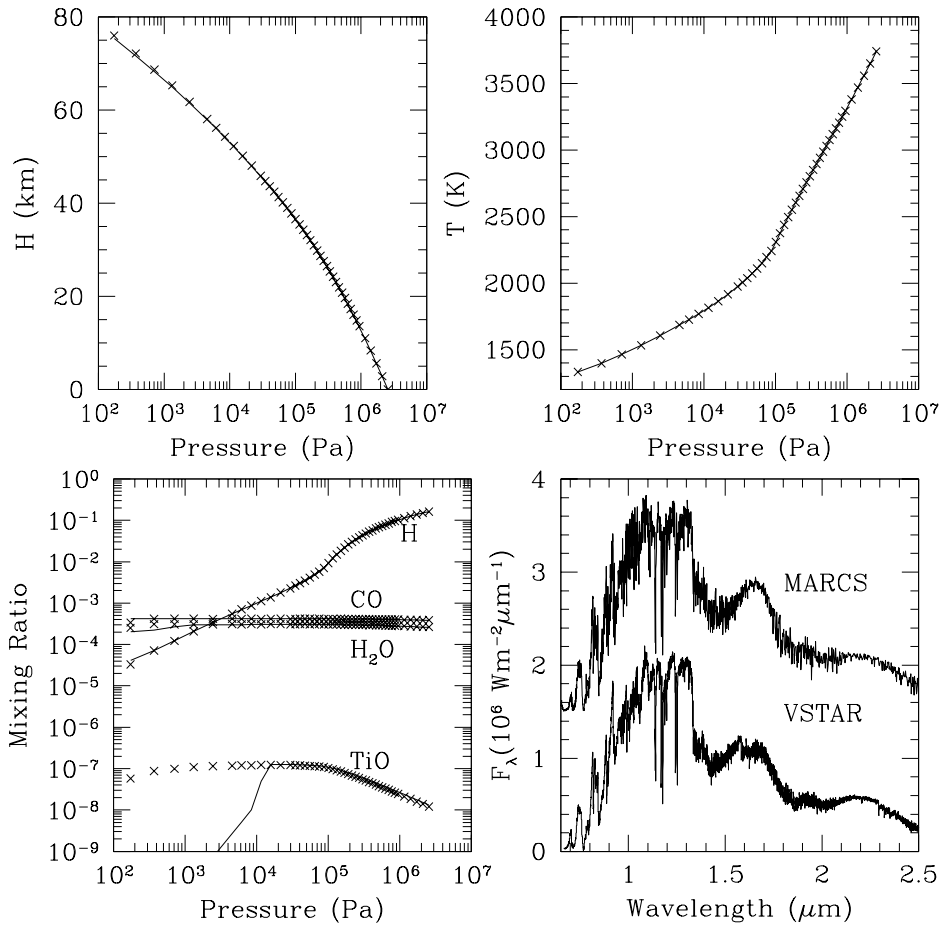}
\caption{Comparison of MARCS and VSTAR for a stellar atmosphere model with 
$T_{\mbox{\scriptsize eff}}$ = 2500 K, $\log{g}$ = 5 and solar metallicity. The height,
temperature and mixing ratios of several important species are plotted as a function of gas
pressure. On these plots crosses are the values from MARCS, and the line shows the values
from VSTAR (only the temperature profile is taken from the MARCS model. All other quantities are
calculated independently by VSTAR and MARCS). The flux spectra from MARCS and VSTAR are
shown in the bottom right panel, with the MARCS data offset by 1.5 $\times 10^{6}$. }
\label{fig_marcs}
\end{figure*}

The MARCS stellar atmosphere code \citep{gustafsson08} is a good choice for a comparison with
VSTAR, since it uses modern abundances and opacities, similar to those used in VSTAR, and has
been used to generate a large grid of model atmospheres with detailed information available on
the model structure. The comparison shown here is for a plane parallel model with 
$T_{\mbox{\scriptsize eff}}$ = 2500 K, g = 1000 m s$^{-2}$ ($\log{g}$ = 5 in c.g.s units), and
solar metallicity. To make the comparison we take the temperature as a function of gas pressure
from the MARCS model, and use this as the input for calculating an equivalent VSTAR model.

Molecular species
included in the VSTAR model were: H$_2$O (BT2), CO, CaH, MgH, FeH, CrH, TiO and VO. Lines of
alkali metals were taken from VALD \citep{piskunov95,kupka99}, and other atomic species were
from the Kurucz line lists.

Figure \ref{fig_marcs} shows some comparisons of data from MARCS and VSTAR. These include the
height above the lowest level modelled (determined from hydrostatic balance), and the mixing
ratios of several important species. On these plots crosses are the values from MARCS and the
lines are the values from VSTAR. In most cases the agreement is excellence. The behaviour for TiO
is however quite different in the cooler layers. This is because the VSTAR chemical model
includes condensate formation, whereas MARCS only includes gas phase chemistry. Small differences
in the CO and H$_2$O mixing ratios arise from the same cause. The difference in TiO does not have
much effect on the strength of the TiO bands in the spectrum as these are mostly formed at deeper
levels.

\begin{figure*}
\includegraphics[width=160mm]{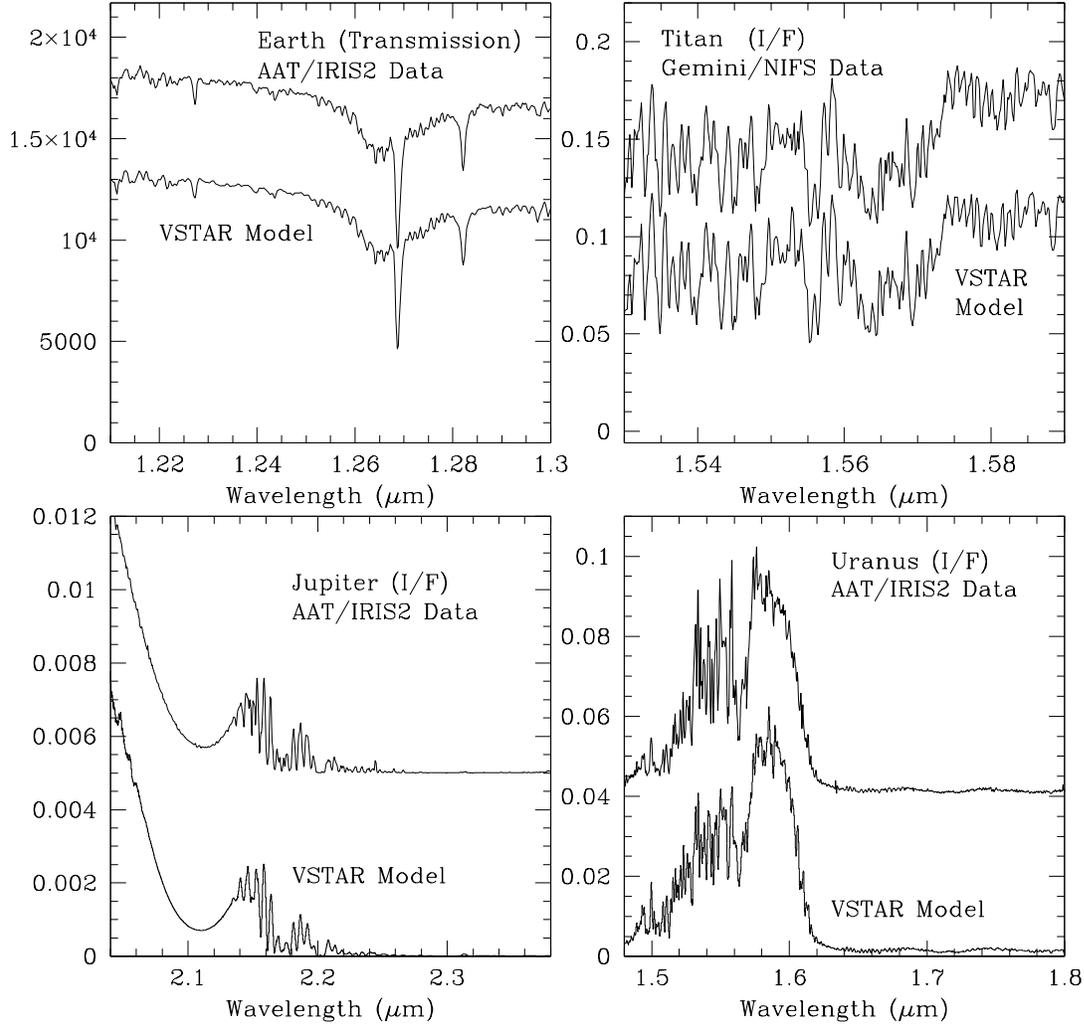}
\caption{Comparison of observed spectrum of solar system objects with VSTAR model spectra. At
top left is shown the spectrum of a G2V star seen through the Earth's atmosphere in the region
of the O$_2$ a-X absorption band, compared with a VSTAR model for a solar spectrum with Earth
atmospheric transmission applied. The other three panels show observed reflectance spectra of Titan,
Jupiter and Uranus compared with VSTAR models. The observed spectrum in each case is offset upwards
relative to the model spectrum.}
\label{fig_ss}
\end{figure*}

The flux spectrum from VSTAR is compared with that from MARCS in the bottom right panel of figure
\ref{fig_marcs}. The fluxes provided with MARCS  are described as "rough estimates of the surface
fluxes" and "are NOT synthetic spectra". Nevertheless it is clear that the values of the fluxes
are very similar and the same absorption fetaures are present in both spectra at similar depths.
A comparison of similar models with observations of an M dwarf spectrum are presented later.

\section{Comparison with Observations}

In this section we present a brief comparison of VSTAR models with observed spectra for a
range of objects. Full details on these and other results will be given elsewhere, but they are
presented here to show the wide range of astronomical objects that can be successfully modelled
using VSTAR. The spectra of solar system objects are from our own observations
with the Anglo-Australian 3.9m telescope at Siding Spring observatory, and its IRIS2
instrument \citep{tinney04}, and from the NIFS instrument \citep{mcgregor03} on the
Gemini North 8m telescope at Mauna Kea Hawaii. The brown dwarf and stellar spectra
are taken from the NASA Infrared Telescope Facility (IRTF) spectral library
\citep{cushing05,rayner09} and are taken with the SpeX instrument on IRTF at a
resolving power of R$\sim$2000.

The spectra considered here are all in the near-infrared spectral region (1 -- 2.5
$\mu$m). For late type dwarfs this is where the flux peaks, and for
spectroscopy of planets, this is a region containing many interesting rovibrational
molecular bands. While this is the wavelength region VSTAR has normally been applied
to, there is nothing inherent in VSTAR that restricts it to this region.
VSTAR should, in principle, be usable from UV to microwave wavelengths.

\subsection{Solar System Objects}

Figure \ref{fig_ss} shows a number of comparisons of observed spectra with VSTAR models 
for solar system atmospheres. The Earth atmosphere transmission comparison (top left panel) is
described in more detail in \citet{bailey07a} and \citet{bailey08b} and uses the observed
spectrum at a resolving power of R ($=\lambda/\Delta\lambda$) $\sim$ 2400 of the G5V star BS 996 with the 3.9m Anglo-Australian Telescope (AAT) and its IRIS2
instrument. This spectrum is compared with a model of a solar spectrum as seen through a
modelled Earth atmosphere transmission spectrum calculated using VSTAR. Of the spectral
features only that near 1.28 $\mu$m is a stellar line. Other features are atmospheric
absorptions of O$_2$, CO$_2$ and H$_2$O, with the strongest feature being the O$_2$ a-X band
at 1.27 $\mu$m and its associated collision induced absorption. The data and model agree to
better than 1\%.

The bottom left panel shows the reflected light spectrum of Jupiter in the near-infrared K
band (2.04 -- 2.38 $\mu$m) also observed with the AAT and IRIS2. The VSTAR model used here
is described in more detail in \citet{chudczer11}. The absoprtion features present in this
spectral region are due to methane, and to the collision induced absoprtion of
H$_2$-H$_2$ (see section \ref{sec_cia}), with the latter producing the broad smooth feature centred on
2.11 $\mu$m. Stratospheric and tropospheric clouds are included in the model with the details 
of the cloud properties and optical depths given in \citet{chudczer11}.

The right hand panels of figure \ref{fig_ss} show spectra of Titan and Uranus in the
spectral region covering the 1.55 $\mu$m methane ``window'' that lies between strong
methane absoprtion bands. Modelling of the spectra of these objects at this
wavelength using line-by-line methods has only become possible very recently with the
availability of the latest laboratory spectral line data for methane as described in
section \ref{sec_ch4}, in particular the data of \citet{wang11}. The Titan spectrum was
obtained with the Near Infrared Integral Field Spectrometer (NIFS) on the Gemini North 8m
telescope, and the data and VSTAR model are described more fully by \citet*{bailey11}. The
absorption lines in this region are mostly weak lines of CH$_4$, but there are also lines
of CH$_3$D and CO. A model for Titan's aerosols based on that of \citet{tomasko08} is also
used in this analysis. The models enabled Titan's D/H ratio and CO abundance to be determined as
described in \citet{bailey11}.

The Uranus comparison in the lower right panel uses data from IRIS2 on the AAT. The VSTAR
model spectrum is calculated from the new low temperature methane line data as described in
section \ref{sec_ch4}. The model includes clouds with a mean particle size of 1 $\mu$m, and with
the main cloud layers being at 2 -- 3 bars and at 6 -- 10 bars, which is similar to cloud
distributions derived by \citet{sromovsky06} and \citet{irwin07}.

While not shown here VSTAR has also been used successfully to model the spectra of other
solar system planets. VSTAR models for the Venus night side are used in the analyses
described by \citet{bailey09} and \citet{cotton11}

\subsection{Brown Dwarf}

\begin{figure}
\includegraphics[width=90mm]{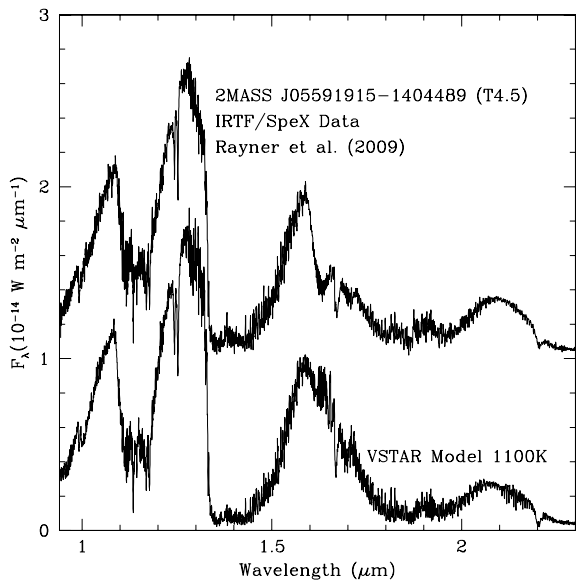}
\caption{Comparison of observed spectrum of the T4.5 dwarf 2MASS J055591915$-$1404489
with a model spectrum
calculated with VSTAR for $T_{\mbox{\scriptsize eff}}$ =
1100K and $\log{g}$ = 5. The observed spectrum is offset from the model upwards by 1
($\times$ 10$^{-14}$)}
\label{fig_bd}
\end{figure}

\begin{figure}
\includegraphics[width=90mm]{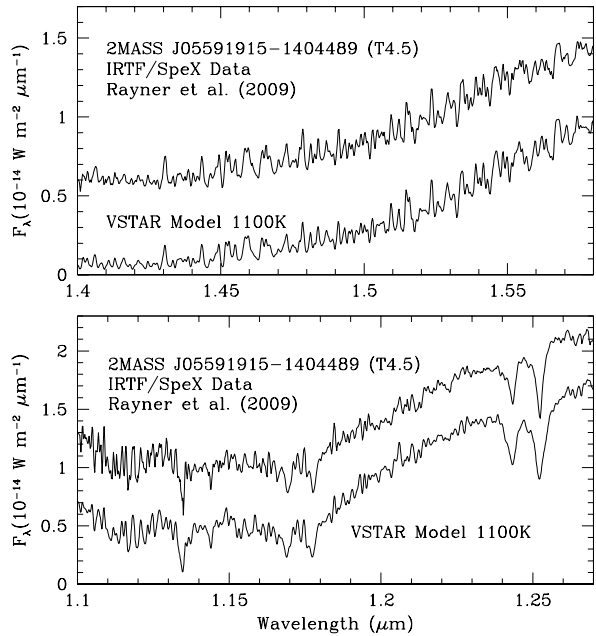}
\caption{Expanded view of two sections of the spectrum shown in figure
\ref{fig_bd}. The observed spectra are offset from the model upwards by 0.5
($\times$ 10$^{-14}$)}
\label{fig_bdh}
\end{figure}

Figure \ref{fig_bd} shows an observed spectrum of the T4.5 dwarf 2MASS
J05591915$-$1404489 taken from the IRTF spectral library \citep*{cushing05,rayner09} 
compared with a VSTAR model spectrum. The VSTAR spectra used for this comparison
are based on brown dwarf
model P-T structures taken from Figures 8 and 9 of \citet{burrows06}. The VSTAR
models included molecular absorption due to H$_2$O (BT2), CO, CH$_4$, CaH, MgH, CrH,
FeH, TiH. Lines of alkali metals are included from VALD \citep{piskunov95,kupka99}.
The far wings of very strong sodium and potassium lines in the visible have a
significant effect on the shape of brown dwarf spectra in the 1 $\mu$m region. We
used a $\chi$ factor model for the far wing shapes of these lines with an exponentially
decrease between 500 and 7500 cm$^{-1}$ from the line centre, with the parameters
adjusted to provide a good match to the data. More physically based models for the
far wing shapes of alkali metal lines are described by \citet{burrows03v} and
\citet{allard03}. 

A number of models were tried with different temperatures and gravities, and best
agreement with the data was found for a model with $T_{\mbox{\scriptsize eff}}$ =
1100 K, g = 1000 m s$^{-2}$ ($\log{g}$ = 5 in c.g.s. units) and solar metallicity.
This is in reasonable agreement with other results for this object.\citet{stephens09} find 
$T_{\mbox{\scriptsize eff}}$ = 1200 K, $\log{g}$ = 4.5 while \citet{delburgo09} find 
$T_{\mbox{\scriptsize eff}}$ = 1002 K, $\log{g}$ = 4.9 using high resolution spectra.

The model spectrum agrees well with the observed spectrum except in the wavelength region from
around 1.6 -- 1.7 $\mu$m. As discused in section \ref{sec_ch4}, the methane line lists currently
available do not include hot bands in this region and therefore underestimate the total
absorption, and incorrectly model the observed structure. This problem exists with all models
for T-dwarf spectra. There is also a discrepancy between the model and observations in the 2.1
$\mu$m region where the model is slightly too low.

In other wavelength regions, where methane absorption is not significant, the agreement between
model and observations is excellent. What might, at first sight, appear to be noise on the
spectra, is in fact spectral structure due to many absorption lines. This is apparent from
figure \ref{fig_bdh} where two wavelength regions (1.1 -- 1.27 $\mu$m and 1.4 -- 1.58 $\mu$m)
are shown on an expanded scale.  

The model used here did not include dust absorption and scattering. This is acceptable for 
objects as late as T4.5. For earlier type brown dwarfs, and in particular the L-dwarf class, the effects of
dust are important. While VSTAR is quite capable of modelling the radiative effects of dust in
such systems, there are currently considerable diferences in the assumptions about dust
properties used in different models. Some models use large ($\sim$100 $\mu$m) dust particles
\citep{burrows06} while others use small (sub-$\mu$m) particles \citep{allard01}. For the
purposes of this paper we have avoided the extra complexity of dusty brown dwarf models.

Our models also assume equilibrium chemistry. Some brown dwarf models include a non-equlibrium
treatment of the chemistry of some species such as CH$_4$, CO and H$_2$O
\citep{saumon03,stephens09} which allows for the effects of vertical mixing in the atmosphere.
While this could be implemented as a modification to our chemical model we have not done so for
the results presented here. The effects on the near-IR spectrum appear to be small
\citep{hubeny07} with more significant effects at 4 -- 14 $\mu$m wavelengths.

\begin{figure}
\includegraphics[width=90mm]{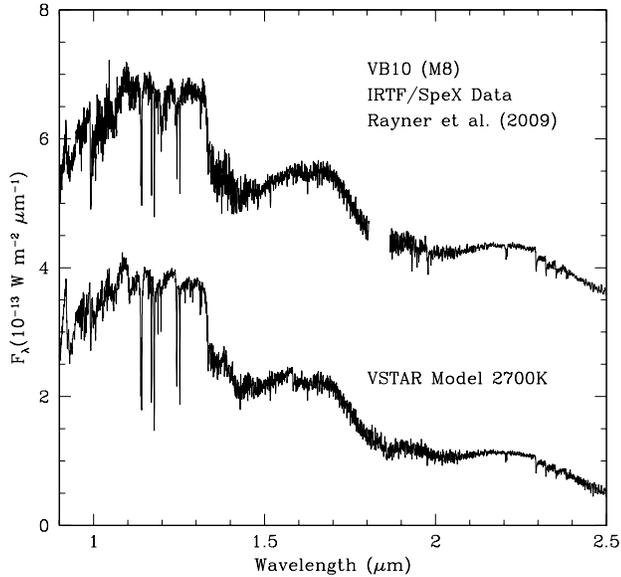}
\caption{Comparison of observed spectrum of the M8 dwarf VB10 with a model spectrum
calculated with VSTAR using a MARCS model structure for $T_{\mbox{\scriptsize eff}}$ =
2700K and $\log{g}$ = 5. The observed spectrum is offset from the model upwards by 3
($\times$ 10$^{-13}$)}
\label{fig_vb10}
\end{figure}

\begin{figure}
\includegraphics[width=90mm]{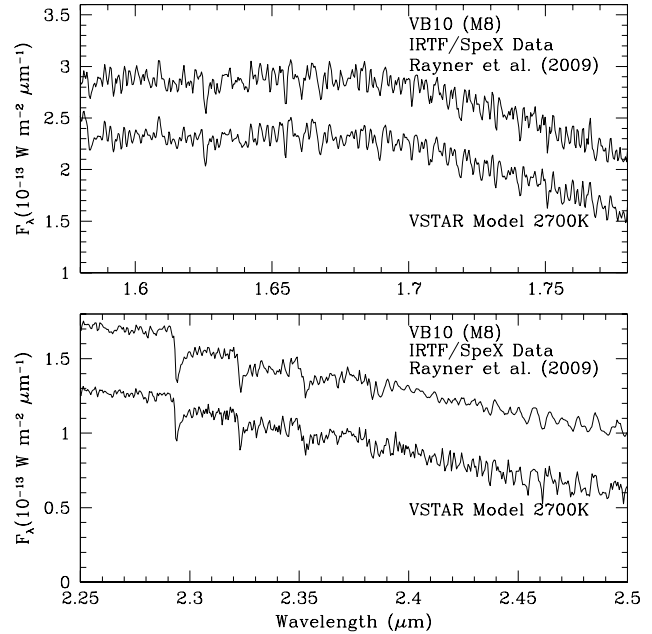}
\caption{Expanded view of two sections of the spectrum shown in figure \ref{fig_vb10}.
The observed spectra are offset from the model upwards by 0.4
($\times$ 10$^{-13}$). The upper panel shows primarily absorption in the E-A band of FeH.
The lower panel shows absorption due to CO and H$_2$O.}
\label{fig_vb10h}. 
\end{figure}

\subsection{M Dwarf}

Figure \ref{fig_vb10} shows an observed spectrum of the M8 dwarf VB10 (Gl 752B) taken
from the IRTF spectral library \citep*{cushing05,rayner09} compared with a model spectrum
calculated with VSTAR. The model structure was taken from the MARCS grid of model
atmospheres \citep{gustafsson08} and is a model for $T_{\mbox{\scriptsize eff}}$ = 2700 K,
g = 1000 m s$^{-2}$ ($\log{g}$ = 5 in c.g.s. units) and solar metallicity. In other respects the
model is similar to that described in section \ref{sec_marcs}. A number of different effective
temperatures were tested with the 2700 K model providing the best match to the observed spectrum.

Figure \ref{fig_vb10h} shows expanded views of two regions of the spectrum. The upper panel
shows the 1.58 -- 1.78 $\mu$m region. This is dominated by absorption lines in the E-A band
of FeH, which are included in our model using the line list of \citet{hargreaves10}. This
band is essential to get a good model of the spectrum in this region. However, including this 
line list results in a small step in the modelled spectrum at 
1.582 $\mu$m (visible in figure \ref{fig_vb10}) which is the shortest wavelength included 
in this list. This step is not seen in the observed spectrum, and suggests that there is more
FeH absorption at shorter wavelengths that should be included. The lower panel of figure
\ref{fig_vb10h} is 2.25 -- 2.5 $\mu$m region showing that the
model does a good job of representing the detailed structure of the CO and H$_2$O bands in this
region. As for the brown dwarf case we have used a dust free model. It is quite possible that dust
is present in such a late type M dwarf, and its inclusion might change our conclusions about the best fitting
effective temperature. 

The upper limit on the layer temperature for modelling stellar atmospheres using VSTAR is 
currently 6000 K set by the use of data from
the JANAF tables in our chemical model. Since a model usually requires a range of layer
temperatures extending to about a factor of two above and below the effective temperature of the
star being modelled, the maximum effective temperature is around 3000 K.

\section{Conclusions}

We have desribed the techniques used in the VSTAR code to calculate model spectra for
objects including solar-system planets, brown dwarfs and M dwarfs. While there are other
codes used for modelling the spectra of these objects, VSTAR has a number of unique
features.

\begin{itemize}

\item It is capable of being used for a very wide range of objects, ranging from the coolest solar
system planets (Uranus with layer temepratures down to 60 K) to stars with layer
temperatures up to 6000 K.

\item It has been tested by comparison with both the Earth atmosphere radiative transfer code
(RFM) and a stellar atmosphere code (MARCS).

\item Not only is a rigorous approach to radiative transfer used, but this has been tested
against radiative transfer benchmarks and shown to be accurate to levels of 10$^{-5}$ to
10$^{-6}$.

\end{itemize}

An obvious application of VSTAR is to the modelling of exoplanet spectra. VSTAR has the
proven ability to model both cool planets in our own solar system, and hotter objects such
as brown dwarfs and M dwarfs, and thus covers the full temperature range expected to be
encountered in exoplanets. The full treatment of the angular dependence of scattering
incorporated in VSTAR means that it can be used to predict the phase dependence of an
exoplanet spectrum around its orbital cycle.

Work is currently in progress on extending VSTAR to include polarized radiative transfer.
The phase variation of polarization can potentially provide important information about an
exoplanet's atmosphere or surface \citep{bailey07,stam08,zugger10}
 
\vspace{0.5cm} 

{\bf \noindent Acknowledgments}

Based on observations obtained at the Gemini Observatory which is operated by the
Association of Universities for Research in Astronomy, Inc. under a cooperative agreement
with the NSF on behalf of the Gemini partnership: the National Science Foundation (United
States), the Science and Technology Facilities Council (United Kingdom), the National
Research Council (Canada), CONICYT (Chile), the Australian Research Council (Australia),
Ministerio da Cienca e Tecnologia (Brazil) and Ministerio de Ciencia, Tecnologia e
Innovacion Productiva (Argentina). The Titan observations were obtained as part of the
system verification of the NIFS instrument under program GN-2006A-SV-128. We thank the
staff of the Australian Astronomical Observatory for assistance in obtaining observations with the
Anglo-Australian Telescope.

\label{lastpage}

\end{document}